\documentclass[iop]{emulateapj}

\usepackage{amssymb, amsmath, natbib, graphicx, multirow}
\usepackage{xspace}
\usepackage{relsize, isotope}
\usepackage{color}
\usepackage[normalem]{ulem}
\usepackage{ulem}

\newcommand{\ten}[1]{10\ensuremath{^{#1}}}
\newcommand{\coo}{CO$_2$\xspace}
\newcommand{\hoh}{H$_2$O\xspace}
\renewcommand{\micron}{$\mu$m\xspace}
\newcommand{\kms}{km~s$^{-1}$\xspace}
\newcommand{\sqcm}{cm\ensuremath{^{-2}}\xspace}

\newcommand{\paperi}{Paper~{\sc I}\xspace}

\shorttitle{}
\shortauthors{}
\slugcomment{Revised \today}

\begin{document}
\title{The Gas-Rich Circumbinary Disk of HR 4049.\\ II: A Detailed Study
of the Near-Infrared Spectrum} 

\author{S.E.~Malek\altaffilmark{1} and J.~Cami\altaffilmark{1,2}}
\altaffiltext{1}{Department of Physics \& Astronomy, University of Western
  Ontario, London, ON N6A 3K7, Canada}
\altaffiltext{2}{SETI Institute, 189 Bernardo Avenue, Suite 100,
  Mountain View, CA 94034, USA}
\email{sarahemalek@gmail.com}
\email{jcami@uwo.ca}

\normalem

\begin{abstract}
HR 4049 is a peculiar evolved binary which is surrounded by a
circumbinary disk. Mid-infrared observations show that the disk is
rich in molecular gas and radially extended.  To study the properties
of this disk, we re-analyzed a set of near-infrared observations at
high spectral resolution obtained with Gemini-Phoenix. These data
cover absorption lines originating from the first overtone of CO and
from \hoh in the 2.3 $\mu$m region as well as more complex
emission-absorption profiles from \hoh and the fundamental mode of CO
near 4.6 $\mu$m.  By using an excitation diagram and from modeling the
spectrum, we find that most of the CO overtone and \hoh absorption
originates from hot gas ($T_{\rm ex} \approx 1000$~K) with high column
densities, consistent with the mid-infrared data.  The strong emission
in the wavelength range of the CO fundamental furthermore suggests
that there is a significant quantity of gas in the inner cavity
of the disk. In addition, there is a much colder component in the line
of sight to the disk. A detailed analysis of the overtone line
profiles reveals variations in the line widths which are consistent
with a radially extended disk in Keplerian rotation with hotter gas
closer to the central star.  We estimate the mass of the primary to be
$\sim 0.34$~M$_\sun$ and discuss the implications for its evolutionary
status.
\end{abstract}


\section{Introduction}

\object{HR 4049} is considered the prototype for a class of evolved
binaries with peculiar properties. The effective temperatures and
luminosities of the members of this class suggest that they are in the
post-asymptotic giant branch (post-AGB) phase of their evolution, but
their evolutionary path is thought to be severely affected by the
presence of a close companion \citep[see][for a
  review]{VanWinckel1995}.

Like the other members of its class, HR~4049 shows a significant
infrared (IR) excess and a time-variable optical and ultraviolet (UV)
deficit \citep{Lamers1986}, which suggests the presence of a massive
circumbinary disk.  In addition, its photosphere shows a severe
depletion in refractory elements, but it has roughly solar abundances
in volatiles.  This unusual depletion pattern is generally attributed
to the formation of dust (incorporating refractory elements) in a
circumbinary disk, followed by the re-accretion of the depleted gas
onto the star \citep{Mathis1992, Waters1992}.

Since the circumbinary disk plays an important role in determining the
properties of HR~4049, it has been studied extensively
\citep[e.g.][]{Waelkens1991a,Dominik2003,Acke2013}.  While the IR
excess is a clear indication of the presence of dust in a stable disk
around the system, it has been difficult to determine the nature of
the dust in the disk. Indeed, it is not even clear whether the dust in
the disk is oxygen-rich or carbon-rich.  While the IR spectrum shows
the clear presence of gas that is typically associated with
oxygen-rich environments (e.g. \coo, \hoh, OH;
\citealt{Cami2001,Hinkle2007}; see also \citealt[][ \paperi
  hereafter]{Malek2013}) there is no trace of corresponding
oxygen-rich dust features such as silicates or oxides. Instead, the
spectral energy distribution (SED) of the disk resembles a 1150~K
black body down to sub-millimeter wavelengths \citep{Dominik2003},
while also showing strong emission features due to polycyclic aromatic
hydrocarbons (PAHs). This PAH emission however does not originate from
the disk, but from what appear to be bipolar lobes \citep{Acke2013}.

The 1150~K black body SED can be reproduced by a ``wall model", in
which the circumbinary disk is vertically extended and the dust is
very optically thick, effectively producing a radiating inner wall at
a temperature of $1150 \pm 150$~K, with a scale height of $\sim$3~AU
at the inner wall distance of 10~AU \citep{Dominik2003}. Because of
the high opacity, the dust beyond the inner rim is cold and does not
contribute to the IR excess. However, the wall model has been
challenged by recent interferometric observations by \citet{Acke2013}
that show a more extended distribution of material emitting strongly
in the IR than the wall model can accommodate. Instead, these authors
suggest emission from optically thin dust with smooth opacity
profiles.

In \paperi, we present a detailed analysis of the molecular emission
in the mid-IR spectrum and conclude that neither of these models is
fully consistent with the properties of the molecular gas (in
particular \coo, \hoh and CO). Indeed, while we find that the gas
originates from a radially extended disk, we also determine that it is
very optically thick across most of the mid-IR. Radiative trapping by
this optically thick gas then results in a warmer and more homogeneous
temperature structure than previously considered for the disk. 

Additional clues to the geometry and properties of the disk can
  be inferred from observations in the near-IR at high spectral
  resolution.  \citet{Hinkle2007} presented such observations,
examining three regions of the near-IR spectrum of HR 4049 at 2.3, 3.0
and 4.6~\micron. They detect many CO, OH and \hoh lines and identify
distinct components in the system. They propose that the absorption in
the CO overtone originates from gas in Keplerian rotation along the
inner rim of the disk in the wall model. Furthermore, they suggest
that the gas is slowly streaming out over the edge of the wall and
over the disk, causing the more complex emission-absorption line
profiles in the 4.6 \micron region. Finally, they found evidence for a
cold gas component. 

Clearly, the near-IR observations contain a lot of information about
the properties, geometry and kinematics of the gas disk. It is thus
important to investigate whether we can reconcile these data with the
disk model inferred from mid-IR observations.  Therefore, we
re-examine the near-IR data presented by \citet{Hinkle2007}.  This
paper is organized as follows. In Section~\ref{sec:observations} we
briefly describe the observational data. We present our analysis
of the spectrum in Section~\ref{sec:analysis}, discuss our results in
Section \ref{sec:discussion} and present our conclusions in Section
\ref{sec:conclusion}.

\section{Observations}
\label{sec:observations}

The data we discuss in this paper are high-resolution ($R\approx
50,000$) observations in the near-IR ($\lambda \sim 1-5 \mu$m) carried
out with the Phoenix spectrograph \citep{Hinkle2003} from the National
Optical Astronomy Observatory (NOAO) mounted on Gemini South. These
observations primarily targeted CO fundamental and overtone lines and
were described and discussed previously by \citet{Hinkle2007}.

A large number of telluric lines are present in the near-IR;
Fig.~\ref{fig:overtone_fits} shows an illustrative atmospheric
radiance spectrum from HITRAN \citep{Rothman20134} on the
web\footnote{http://hitran.iao.ru/}. Additional observations of a hot
star (without stellar lines in this wavelength range) at the same
airmass as HR~4049 were used to divide out these lines. This telluric
correction is generally adequate; however, some residuals remain due
to imperfect cancellations, especially in wavelength ranges where
there are a lot of lines (e.g. near 2.319 $\mu$m; see
Fig.~\ref{fig:overtone_fits}). In most cases though, the residuals are
significantly smaller than the depth of the lines we study here. The
telluric lines in the hot star spectrum were also used to achieve a
very accurate wavelength calibration, with residuals of typically
0.25~\kms. For more details about these observations or the data
reduction aspects, we refer to \citet{Hinkle2007}. 

For each individual data segment, we then performed a heliocentric
velocity correction and normalized the data to the continuum. Next, we
used a weighted mean \citep[adopting the signal-to-noise, S/N, ratios
  provided by][]{Hinkle2007} to merge all the individual segments into
a single, final spectrum. Finally, we rebinned the resulting spectrum
onto a constant resolution wavelength grid for later comparison to our
model spectra.

\section{Analysis}
\label{sec:analysis}

We begin our analysis of these data with the CO overtone lines in the
2.3 \micron region.  There are several good reasons for this.  First,
in this range, the spectrum reveals simple, pure absorption profiles
with relatively little blending between the lines. Since they are
intriniscally much weaker, these lines will furthermore have a much
lower optical depth than the fundamental lines for a given column
density, thus offering the best prospects to reliably determine the
temperatures and column densities. Finally, since the gas in this
region of the spectrum is absorbing continuum radiation, its location
is constrained: it must be directly along the line of sight to the
dusty disk and/or the central star; note however that the stellar
continuum will only contribute $\sim$25\% of the flux at this
wavelength. We will then use our results from the 2.3 \micron range to
better interpret the emission-absorption spectrum at 4.6 \micron.

\subsection{The 2.3 \micron Region}
\label{sec:analysis2}

\begin{figure*}[t!]
\centering
\includegraphics[width=1\textwidth]{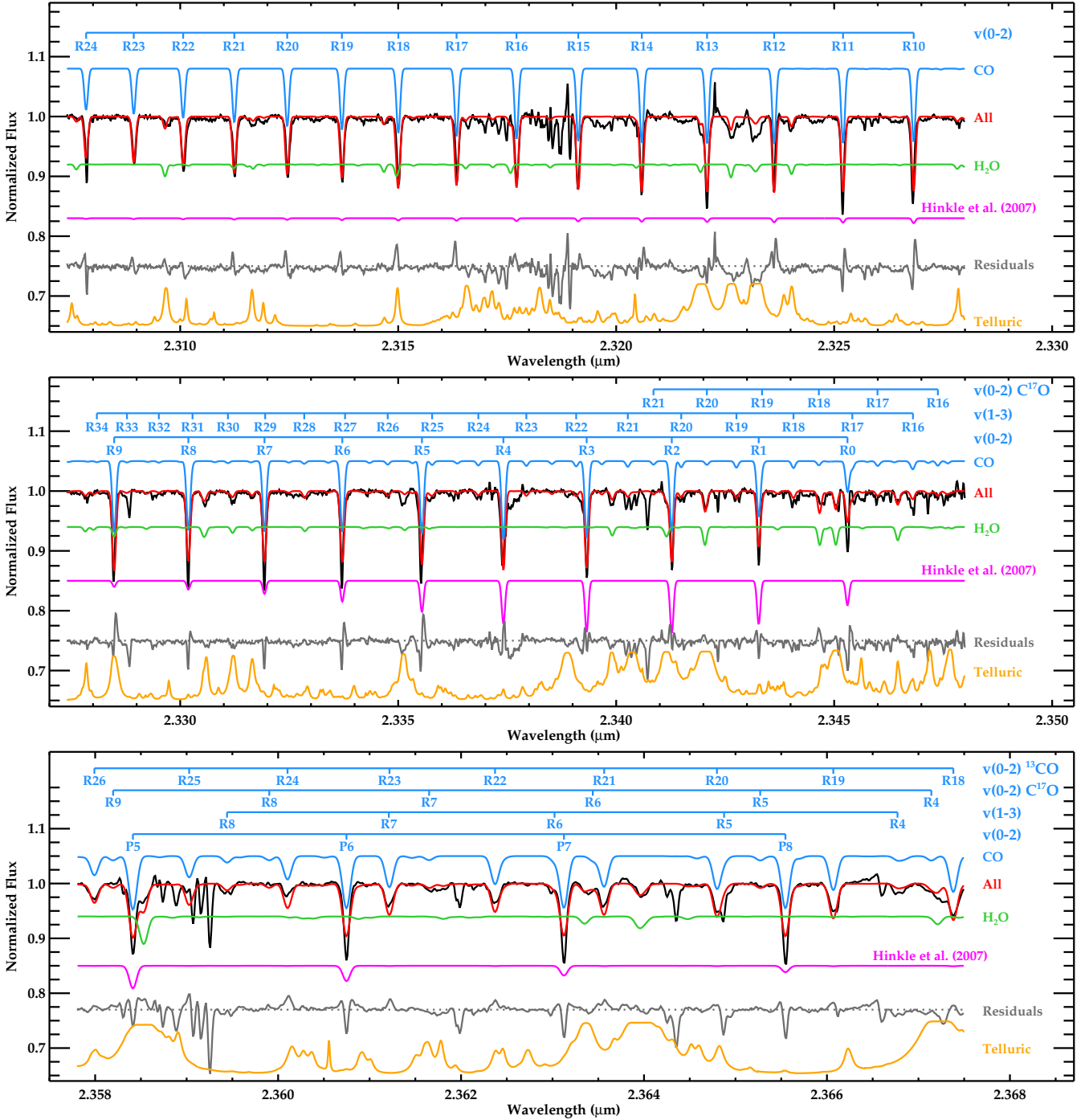}
\caption{The data for the overtone CO lines (black) and our best fit
  model for CO and water (red). Residuals are shown in grey. The
  individual contributions of CO and H$_2$O are shown respectively
  above (blue) and below (green) the data. We have labeled individual
  $v(2-0)$ and $v(3-1)$ transitions for the main isotope, and for
  $^{13}$CO and C$^{17}$O. For comparison, we also show a model using
  the temperatures and column densities determined by
  \citet[][magenta]{Hinkle2007}, and an illustrative telluric radiance
  spectrum (orange) from HITRAN on the web. Note that the wavelength
  scale of the bottom panel is different from the other two panels.}
\label{fig:overtone_fits}
\end{figure*}

The combined spectrum of HR 4049 in the 2.3 \micron range is shown in
Fig.~\ref{fig:overtone_fits}. We remind the reader that
\citet{Hinkle2007} used these data to produce an excitation diagram
for CO, and found that the absorption originates from two distinct
layers of CO gas in local thermodynamic equilibrium (LTE): a warm
layer ($T = 530$~K, $N = 1.65 \times$ \ten{17} \sqcm) and a cold one
($T = 40$~K, $N = 4.58 \times$ \ten{18} \sqcm).

We first used these parameters to create a plane-parallel slab model
in which 1200~K black body radiation (representing the dust) is
absorbed by the warm layer of CO gas which is in turn absorbed by the
cold layer of CO gas. We used the CO line list from
\citet{Goorvitch1994} and adopted a Gaussian line profile with a width
of 10~\kms. This corresponds to the measured full width at half
maximum (FWHM) for the overtone lines; note that \citet{Hinkle2007}
suggest 16 \kms; however, this results in lines that are too broad
(see also Sect.~\ref{sec:discussion_kinematics}). When we compared
this model to the observations (see Fig.~\ref{fig:overtone_fits}), we
found that the model reproduces neither the depths nor the relative
strength ratios of the individual rovibrational lines. This
discrepancy could be due to the use of a different line list; at any
rate, it warrants an independent analysis.

\paragraph{Excitation Diagram}
First, we constructed our own excitation diagram for the CO first
overtone $v(0-2)$ lines. We fit a Gaussian to each CO line and
integrated these to determine the equivalent width ($W_\lambda$).
Then we determined the population of each level using

\begin{equation}
N = 1.13 \times 10^{20} W_\lambda / (f_{J' J''} \lambda^2) \,
\end{equation}

\noindent where $N$ is the column density (in cm$^{-2}$), $f_{J' J''}$
is the oscillator strength for a rovibrational transition
\citep[obtained from the CO line list by][]{Goorvitch1994}, $\lambda$ is the
central wavelength of the transition and both $W_\lambda$ and
$\lambda$ are measured in \AA~\citep{Spitzer1978}. 

\begin{figure}[t!]
\includegraphics[width=0.5\textwidth]{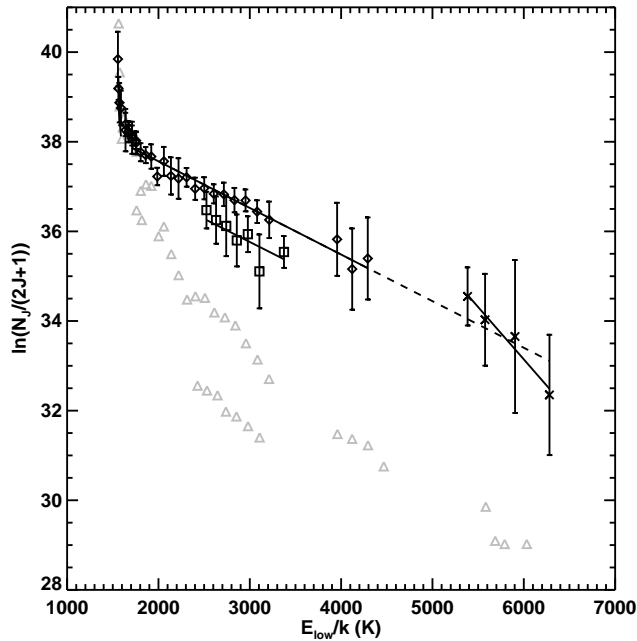}
\caption{Excitation diagram for the overtone CO lines with the
  $v(0-2)$ lines indicated by diamonds, the $v(1-3)$ points indicated by
  crosses and the $^{13}$CO points indicated by squares.  The gray
  triangles indicate the points on the excitation diagram presented by
  \citet{Hinkle2007}.  The solid black lines indicate the best fits to the
  different components. The dashed line shows the extrapolation of the
  lines originating from the ground state to those from the first
  vibrationally   excited level.} 
\label{fig:co_excitation}
\end{figure}

We present the resulting excitation diagram in
Fig.~\ref{fig:co_excitation}. Qualitatively, our results are
similar to those obtained by \citet{Hinkle2007}, in the sense that our
excitation diagram is consistent with two layers of optically thin CO
in LTE---a warm and a cold one. However, we find very different values for the
temperatures and column densities of the CO gas: our ``hot'' layer has
a temperature of 970 $\pm$ 40 K and a column density of (1.19 $\pm$
0.04)$\times 10^{19}$ cm$^{-2}$.  The temperature we determine is
nearly twice that found by \citet{Hinkle2007} and the column density
is two orders of magnitude larger.  For the ``cold'' layer, we find a
temperature of 40 $\pm$ 10 K and a column density of (6.1 $\pm$
0.5)$\times 10^{18}$ cm$^{-2}$.  Here, the temperature and column
density are relatively similar to those determined by
\citet{Hinkle2007}.

If we extrapolate the hot component to higher energies, we find that
it also fits the overtone lines originating from the first
vibrationally excited state (i.e. the $v(1-3)$ lines) relatively
well. This suggests that the CO may be in---or close to---vibrational
LTE. For the $^{13}$CO lines, we find a column density of (5.5 $\pm$
0.6)$\times$ \ten{18} \sqcm.  This yields a $^{12}$C/$^{13}$C ratio of
2.1 $\pm$ 0.9, which is a little lower the value of 6$^{+9}_{-4}$
reported by \citet{Hinkle2007}.

\paragraph{Model Spectra}
As an independent check, we modeled the CO absorption at 2.3~\micron
using the same methods employed to build the SpectraFactory database
\citep{Cami2010a}.  For these models, we began once again with CO line
lists from \citet[including the $^{13}$CO, C$^{18}$O and C$^{17}$O
  isotopologues]{Goorvitch1994}. From these, we calculated optical
depth profiles assuming a population in LTE and a Gaussian intrinsic
line profile with a width of 0.5 \kms \citep[consistent with the $b$
  value derived by][]{Hinkle2007}. Note though that since the lines
are optically thin, the precise value of the intrinsic line width in
our approach does not matter much as long as the line width is much
smaller than a resolution element.

We tested a few different model configurations. Initially, we
attempted to fit a model consisting of a single layer of LTE gas
absorbing a 1200 K black body, however, we found that this model did
not reproduce all of the CO absorption and found strong residuals,
especially for the low $J$ lines. Thus, it seems that at least two
layers of CO gas were required; this is consistent with the results
from the excitation diagrams. We thus modified our model to
include two layers of gas in a slab geometry absorbing a background
dust continuum.

In our fit to the CO absorption, we used gas temperatures between 50
and 1000~K in increments of 100 K and column densities between
10$^{16}$ and 10$^{20}$ cm$^{-2}$ in increments of log$N$ = 0.2 for
each of the gas layers.  We varied
log[\isotope[12]{C}/\isotope[13]{C}] from 0 to 2;
log[\isotope[16]{O}/\isotope[17]{O}] and
log[\isotope[16]{O}/\isotope[18]{O}] from 0 to 3 in increments of
0.2. We also varied the radial velocity ($v_{\rm rad}$) between -30
and -40 km~s$^{-1}$ in increments of 1 km~s$^{-1}$ and smoothed our
line profiles to yield lines with widths between 2 and 20 km~s$^{-1}$
in increments of 2 km~s$^{-1}$ to represent the observed line
broadening (due to e.g. the rotation of material in a disk).  We
compared each model to the entire 2.3 $\mu$m spectral region and
calculated $\chi^2_\nu$, the reduced $\chi^2$ statistic for each
model.

In Table \ref{table:nirfits}, we present the parameters from our best
fit model and in Fig.~\ref{fig:overtone_fits}, we compare this model
to the spectrum. We note that this model reproduces the CO absorption
very well. In addition, the resulting best-fit temperatures (50~K and
900~K) agree well with those we determined from our excitation diagram
(40~K and 970~K) and also the column densities are comparable (log$N =
18.20$ versus log$N = 18.7$ for the cold layer and log$N = 19.20$
versus log$N = 19.08$ in the hot); this corresponds to a maximum
optical depth of $\tau_{\rm max}\approx 0.5$.  We find a
$\chi^2_\nu$ of 3.54 for our best model however, indicating
significant residuals.

\begin{deluxetable}{l c c}
\tablecolumns{3}
\tablewidth{0pt}
  \tablecaption{Temperatures, column densities, isotope ratios and
   velocity parameters for the model fits to our data (with nominal
   3$\sigma$ uncertainties) from our excitation diagram and the
   isolated CO lines.}
\startdata
\hline \hline
	& Model	& Excitation Diagram \\  \hline \hline
\sidehead{Gas Layer 1}  
Temperature (K) & 900 $\pm$ 50   & 970 $\pm$ 40 \\ [1.1ex]
Log $N$ (CO)    & 19.2 $\pm$ 0.1 & 19.08 $\pm$ 0.01 \\ [1.1ex]
Log $N$ (\hoh) 	& 19.4 $\pm$ 0.1 & $\cdots$ \\ [1.1ex]
\sidehead{Gas Layer 2} 
Temperature (K) & 50 $\pm$ 25    & 40 $\pm$ 10 \\ [1.1ex]
Log $N$ (CO)    & 18.2 $\pm$ 0.1 & 18.7$^{+0.2}_{-0.3}$ \\ [1.1ex]
\sidehead{Overall}
$^{12}$C$/^{13}$C & 1.6$^{+0.4}_{-0.3}$ & 2.1 $\pm$ 0.9 \\ [1.1ex]
$^{16}$O$/^{18}$O & 16$^{+9}_{-5}$ & $\cdots$ \\ [1.1ex]
$^{16}$O$/^{17}$O & 16$^{+4}_{-3}$ & $\cdots$ \\ [1.1ex] \hline \hline 
	& 	& Velocity Analysis \\ \hline \hline
$v_{\rm rad}$ (km s$^{-1}$) & -33.0 $\pm$ 0.5 & -32.7 $\pm$ 0.2 \\ 
$v_{\rm FWHM}$ (km s$^{-1}$) & 10 $\pm$ 1     & 11.0 $\pm$ 0.5 \\
\enddata
\label{table:nirfits}
\end{deluxetable}

Note that our model does not only reproduce the ground state $v(0-2)$
transitions, but also the $v(1-3)$ lines (see e.g.~the R18, R24 and
R28 lines in Fig.~\ref{fig:overtone_fits}), again suggesting that the
CO gas is near vibrational LTE. Note however that the critical density
for the vibrational levels of CO is $n({\rm H}_2) \approx
\ten{12}-\ten{13}$~cm$^3$ \citep{Najita1996}, which is much higher
than any realistic estimate of the density in this
environment. However, as described in \paperi, the disk is extremely
gas rich and significant radiative trapping is occurring. The gas may
thus be radiatively thermalized.

\subsubsection{\hoh}

The residual spectrum still exhibits several additional absorption
features; some of these are telluric (see
Fig.~\ref{fig:overtone_fits}), however, others are consistent with
\hoh at the radial velocity of HR~4049. These features are much
shallower than the CO lines which makes them much less obvious.  We
thus recalculated our model for this region with the addition of \hoh
using the line list from \citet[including the H$_2^{17}$O and
  H$_2^{18}$O isotopologues]{Partridge1997}.

We began with the assumption that \hoh was in the same layers as the
CO (keeping the same layers reduced the number of free parameters) and
varied the column density in each layer between 10$^{16}$ and
10$^{20}$ cm$^2$. The best fit for these models indicated an absence
of \hoh in the cold molecular layer. Given the temperature of this
cold layer, water may not exist in the gas phase here. Thus, we
removed \hoh from our cold CO layer and only included it in the hot
layer. Like the hot CO, we find a high column density for \hoh in this
region of the spectrum (log $N$ = 19.4), but still corresponding to
optically thin lines ($\tau_{\rm max}\approx 0.14$).

\subsection{The 4.6 \micron Region}

\begin{figure*}[t!]
\centering
\includegraphics[width=1\textwidth]{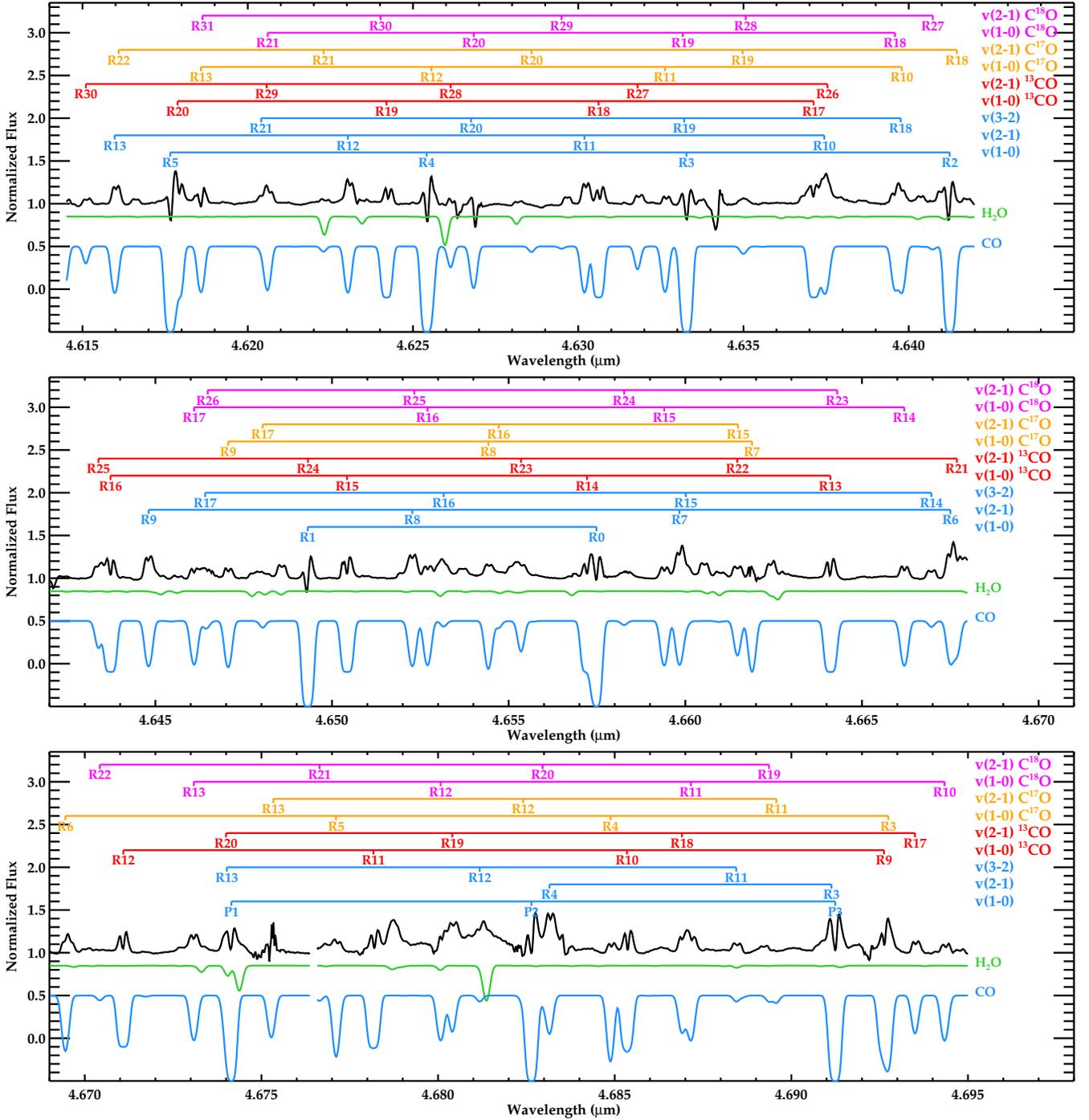}
\caption{\label{fig:co_fundamental}The 4.6 \micron region of the
  spectrum of HR 4049 (black). Model predictions using the best-fit
  parameters for the overtone regions are shown below the data for
  water (green) and CO (blue). The wavelengths of the fundamental
  transitions (ground state and hot bands) are marked and labeled for
  the main CO isotopologue (blue), for $^{13}$CO (red), C$^{17}$O
  (orange) and C$^{18}$O (magenta).}
\end{figure*}

Armed with a good characterization of the absorbing gas in the 2.3
\micron region, we now turn our attention to the lines near 4.6
\micron. We show the spectrum of HR 4049 in this range in
Fig.~\ref{fig:co_fundamental}. In contrast to the pure absorption
lines in the CO overtone region, the spectrum here is dominated by
emission bands (of both CO and \hoh), many of which show a
superposed absorption component as well. Note that some of these
absorption features are not just self-absorption of the emission
lines, but actually absorb continuum radiation as well. 

Such a spectrum, dominated by emission, is somewhat surprising at
first sight given our results from above. The overtone lines at 2.3
\micron revealed the presence of a hot gas in front of the continuum
emission from the star and dust disk. All other things being equal,
the same gas should also produce absorption in the fundamental lines
since they originate from the same (ground) state. In fact, since
the fundamental lines are intrinsically $\sim$100 times
stronger than the overtone
lines and since the $\sim$1200~K black body dust emission is almost
twice as strong at 4.6 \micron than it is at 2.3 \micron, we would
expect much stronger, saturated absorption lines, as we show in
Fig.~\ref{fig:co_fundamental}. Using the same parameters as for the
2.3 \micron region, we find that in this range, both the CO and \hoh
lines would be very optically thick ($\tau_{\rm MAX} \sim 100$ for
CO and $\tau_{\rm MAX} \sim 4$ for \hoh) and thus appear as broad,
saturated lines.  This is in stark contrast with the observations. 

Thus, the emission must originate from gas that is not residing in the
same line of sight as the gas detected in the overtone and must cover
a much larger area to completely fill in the expected
absorption. \citet{Hinkle2007} interpreted this in the framework of
the wall model and suggested that the gas in absorption was along the
inner wall of the disk, while the gas in emission originated from
above the cold dust surface. However, this is not a viable explanation
anymore given our current understanding of the disk from
interferometric and mid-IR spectroscopic observations: the disk is
radially extended, and is warm over a large radial distance
\citep[\paperi]{Acke2013}. Furthermore, the gas emitting in the CO
fundamental must be quite warm as well, as evidenced by the high $J$
lines and the presence of $v(2-1)$ and $v(3-2)$ transitions.

One way to explain these observations is by considering that the
line of sight that we probe at 2.3 \micron is such that the CO is
not absorbing the dust emission, but rather the stellar
continuum. Given our line depths at 2.3 \micron and that
about 25\% of the flux at those wavelengths originate from the star,
this could certainly not be ruled out. The total flux we observe is
then the sum of this absorbing line of sight and the total emission
(gas + dust) of the disk. At the gas temperatures we determined, the
gas in the disk would not emit much at 2.3 \micron, thus the
main effect would be veiling of the absorption lines by the dust. At
4.6 \micron however, the starlight contributes only 5\% to the
continuum, so the depths of the absorption lines would be much
more reduced due to veiling. At the same time, the $\sim$1000~K gas
would also emit very efficiently at 4.6 \micron and any
remaining absorption would be more easily filled in by emission from
hot gas. 

However, this scenario does not work. If the CO gas in the overtone
lines would be absorbing only the stellar radiation, our line of sight
through the disk would only probe material at a zero radial velocity
since any gas motions would be perpendicular to the line of sight. A
detailed analysis of the overtone line profiles (see
Sect.~\ref{sec:discussion_kinematics}) shows that the higher $J$ lines
have much broader line profiles and thus at least a fraction of the
absorption originates from gas with a significant radial velocity;
thus, this gas must be absorbing non-stellar radiation.

Therefore, we must conclude that the gas in the overtone region is
absorbing dust emission as well. To explain the emission in the
fundamental region, there must then be a significant amount of hot
gas located in lines of sight that do not intersect the dust
emission. At this point, there only seem two plausible locations for
that gas.

First, the gas could originate from above
(and/or below) the disk mid-plane, while most of the dust providing
the continuum emission would be located near the mid-plane. As
noted by \cite{Dominik2003}, the dust settling time for a gas-rich
disk is short---typically 150 years for the case of HR 4049; thus, the
dust distribution is most likely concentrated near the mid-plane. At
the same time, if the disk is in hydrostatic equilibrium, the high gas
temperatures imply that the disk must still have a large scale height.
It is thus possible that the dust emission is predominantly
originating from the mid-plane while most of the gas emission is
coming from above the mid-plane. However, there must still be enough
(small) dust grains mixed in with the gas at appreciable vertical
distances from the mid-plane to explain the phase-dependent extinction
\citep{Waelkens1991a,Acke2013}. Furthermore, if the gas is being
heated by the dust, the dust must be able to intercept a sufficient
amount of stellar radiation, which would not be possible if all the
dust is located at the mid-plane. 

A second possibility is that there is hot gas located inside the dust
disk, in the dust-free central cavity. Since the inner boundary of the
disk is determined by the sublimation temperature of the dust and
since the dissociation temperature for CO (and \hoh) is much larger
than the dust sublimation temperature, some gas must indeed be located
inside the disk. Some support for an origin inside the disk can be
found in the larger widths of the emission components compared to the
absorption (by hot gas) in the overtone (see
Sect.~\ref{sec:discussion_kinematics}), pointing to larger velocities
for the emitting component than for the absorbing gas.

\medskip

Finally, we should also briefly discuss the absorption components
in the emission lines. While most emission lines show some evidence
for an absorption component, it is most pronounced in those lines
involving the ground vibrational state. Furthermore, the low $J$
lines show the strongest absorption, including absorption of the
continuum radiation. 
Thus, this absorption component must originate
from a much cooler component (such as the cold component we observe
in absorption at 2.3~\micron) and cannot be due to self-absorption
by the gas (which would not result in continuum absorption). Since
some of the absorption lines are slightly blue shifted compared to
the emission component, this cold gas may be in an outflow, possibly
related to the bipolar lobes. 

\subsection{Kinematics}
\label{sec:discussion_kinematics}

We also investigate the kinematic properties of the gas.  As noted in
Sect.~\ref{sec:analysis2}, the lines at 2.3~\micron appear narrower
than the velocity width reported by \citet{Hinkle2007}. When we fitted
all the overtone lines from the ground state with Gaussians, we
determined that the average full-width at half maximum velocity was
$v_{\rm FWHM} = 10$~\kms rather than their $v_{\rm FWHM} =
16$~\kms. As an additional check, we also incorporated the line width
as a parameter in our full model fit which confirmed this result. Note
that this also agrees with \citet{Lambert1988}, who determined that
the low $J$ lines were unresolved when observed with an instrumental
resolution of 10 \kms.

\begin{figure}[t]
\centering
\includegraphics[width=0.5\textwidth]{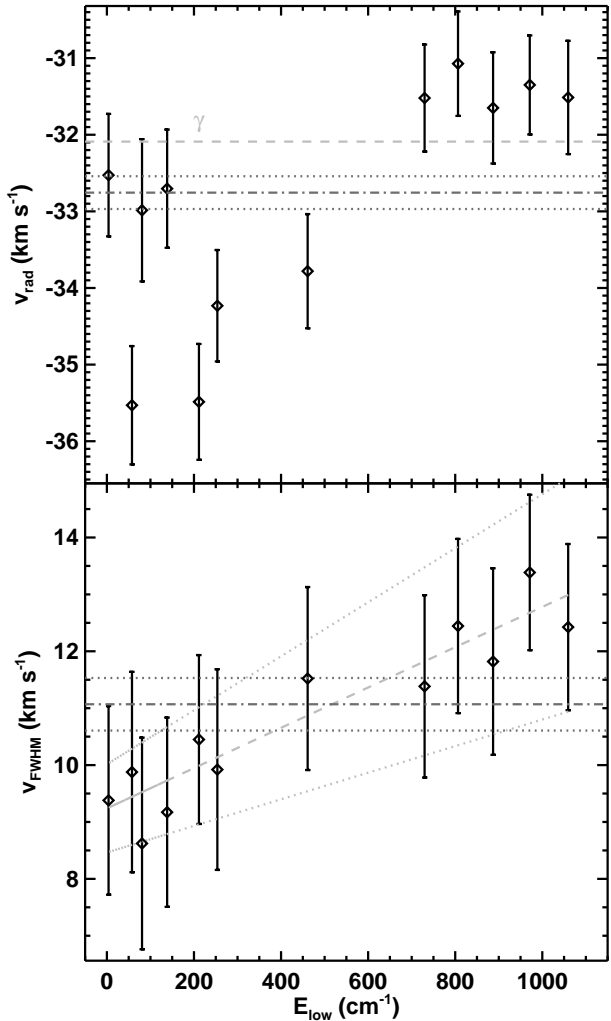}
\caption{The top plot compares the radial velocity of each isolated CO
overtone transition to the lower energy of the transition (with the
zero-point vibrational energy set to 0). The blue dot-dashed line shows
the weighted mean and the dotted lines indicate the range. The red
dashed line indicates the system velocity \citep[$\gamma =
-32.07$~\kms][]{Bakker1998}. The lower plot compares the $v_{\rm
FWHM}$ as a function of lower energy. The blue dot-dashed horizontal line
indicates the weighted mean of all the $v_{\rm FWHM}$ measurements and the blue
dotted lines indicate the uncertainties on this measurement. The red
dashed line indicates a linear fit 
between these parameters, with the dotted red lines showing the
uncertainties in the fit. Error bars on the velocity measurements are
determined with a Monte Carlo technique.} 
\label{fig:overtone_v}
\end{figure}

To further examine the kinematics of the gas in the disk, we fit
unconstrained Gaussian profiles to only the isolated CO overtone
transitions from the ground vibrational state (i.e.~those that are not
blended with lines from hot bands, other isotopologues or \hoh lines)
and examined the resulting radial velocities and line widths. We find
that on average, the lines have radial velocities of -32.7 $\pm$ 0.2
\kms and a range between -35.5 and -31.1 \kms. We also find an average
line width of 11.0 $\pm$ 0.5 \kms and a range between 8.5 and 13.4
\kms. Interestingly, both the radial velocity and width measurements
show some clear variations as a function of the lower state energy of
the lines involved (see Fig.~\ref{fig:overtone_v}). 

\paragraph{Radial Velocities} 
In Fig.~\ref{fig:overtone_v}, we recover a trend previously noted
by \citet[][their Fig.~10, but there for the absorption
component in the {\em fundamental} lines]{Hinkle2007}: the low $J$ lines
are blue shifted compared to the higher $J$ lines which are closer to
the system velocity. This was attributed to a slow outflow
of gas from the system. 

\begin{figure}[t!]
\centering
\includegraphics[width=0.48\textwidth]{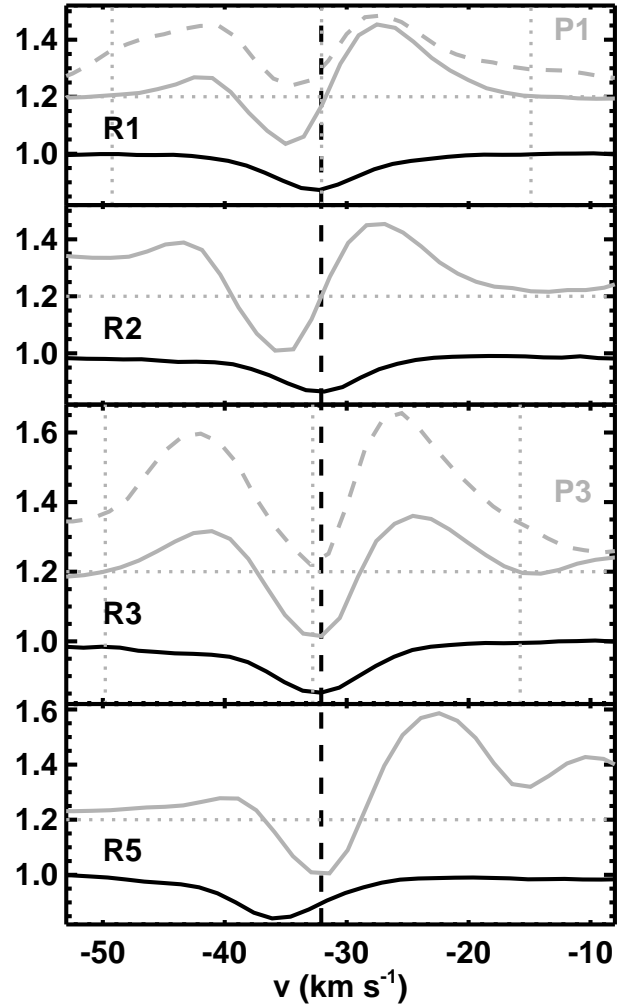}
\caption{A comparison between overtone absorption lines (in black) and
the fundamental emission-absorption lines for which the absorption
component originates from the same vibrational and rotational states
offset by 0.2 (in gray). The R-branch transitions are solid lines
while the P-branch transitions are dashed lines. For the R1 and R3 lines we
indicate the full width at zero intensity and the midpoint in vertical
dotted gray lines. The dashed black line indicates the system
velocity. The horizontal gray dotted line indicates the offset in the
fundamental observations. }
\label{fig:fundamental_v_overtone}
\end{figure}

To compare both absorption characteristics, we show the profiles of the overtone and
fundamental lines that exhibit absorption for lines originating from
the same level in Fig.~\ref{fig:fundamental_v_overtone}. Again, we
selected only the lines which had relatively clean absorption profiles
with as little blending as possible in both the fundamental and the
overtone.

It is clear that the centers of the absorption lines do not correspond
well to each other; the $J=3$ lines are the only ones with a clear overlap. In
addition, the variations in the fundamental lines are different from the
overtone: for the overtone, the low $J$ lines are at the system
velocity and the largest deviations from the system velocity are found
for the R5 and R10 lines; for the fundamental, it is only the lowest $J$
lines that are blue shifted. 

These observations are hard to explain consistently. For
instance, the absorption in the fundamental could on first sight be
explained by a cold outflow from the system. Indeed, with the
deepest absorption (the R2 line) originating from the $J=2$ level,
this blue shifted gas must have $22~ {\rm K} \la T_{\rm ex} \la 50~
{\rm K}$. However, at such temperatures, there would still be a
significant population in the $J=3$ state and we would expect
to also see the same blue shifted absorption quite strongly in the R3
line---but this is not the case. That points to a low-density gas
that is not even in rotational LTE and may be more characteristic
of interstellar gas than circumstellar gas.

\medskip

In the overtone lines, the case for a systematic effect is even less
clear (see Fig.~\ref{fig:overtone_v}), with some low $J$ lines at the
system velocity and others blue shifted. It is thus not clear whether
this represents a true effect (e.g. we underestimate blending by other
lines) or rather points to instrumental or calibration issues.

For completeness, we note that these shifts are only seen in the CO
fundamental of the main isotopologue; there are no discernible trends
in the radial velocities of the absorption components of the hot bands
and other isotopologues of CO.

\paragraph{Line Widths}

An interesting trend becomes obvious when looking at the widths
of the lines as a function of the lower state energy (bottom panel
of Fig.~\ref{fig:overtone_v}): overtone absorption lines originating
from higher energies are broader than the lower energy lines. A
linear fit yields a slope of $(3.4 \pm 1.2) \times 10^{-3}$ km
s$^{-1}$ per cm$^{-1}$. Since only hot gas can contribute to the
lines at higher energies and since cold gas dominates the lower
energy lines, we conclude that the hotter gas has a larger
rotational velocity than the cooler gas, consistent with Keplerian
motion in a disk that is hotter on the inside and cooler on the
outside. Note that thermal broadening is much smaller than the
effect we measure here (at $T \approx 1000$~K, $b \approx 0.75$
\kms). We will discuss this effect further below. 

We also note that in Fig.~\ref{fig:fundamental_v_overtone}, the
emission component is clearly broader than the overtone absorption.
If the broadening mechanism is rotation in a Keplerian disk and the
overtone absorption begins at the interior of the dust disk, this once
more suggests that the emission component originates on the interior
of the dust disk.

\section{Discussion}
\label{sec:discussion}

It is clear from our analysis above that the CO (and \hoh) gas
probed in the near-IR traces a very complex, gas-rich
environment. As pointed out by \citet{Dominik2003}, hydrostatic gas
pressure in a gas-rich disk will not only puff up the disk in the
vertical direction, but it will also act in radial direction and
tend to spread out the disk. Indeed, also the near-IR observations
point to such a radially extended disk, in agreement with our
conclusions from the mid-IR observations in \paperi. 

\subsection{Dust Temperatures}

As we suggest in Sect.~\ref{sec:discussion_kinematics}, the gas
appears to have some temperature variation in which the gas closest to
the binary system is hotter than the gas further away. If the gas and
dust are mixed in, then this should also be true of the dust. 

If the dust were gray and in equilibrium with the stellar radiation
field, the temperature in the disk would vary from 520~K to 370~K at
15 and 30~AU respectively \citep[using a stellar temperature of 7500~K and a
stellar radius of 31 R$_\sun$ as described by][]{Acke2013} or 640 and
450 at 10 and 20~AU. 
The dust must be hotter than this to allow the $\sim1000$~K gas
component to appear in absorption; indeed, it must be hotter than the
gas for the gas to appear in absorption. 

In \paperi, we suggest that the optically thick gas in the disk traps
and re-radiates the stellar radiation, thus the temperatures of the gas and
dust in the disk will increase significantly compared to equilibrium
values and they will become more homogeneous. The question thus remains: what
type of temperature distribution can the disk have while still
reproducing the SED we observe in HR~4049?

As described by \citet{Dominik2003}, the SED can be reproduced by the
sum of equally weighted black bodies with temperatures between 880 and
1325~K. As well, when they assigned weights for the black bodies based on a
power law distribution, they found that the temperatures could vary
between 730 and 1238~K. In general, the maximum dust temperature
possible is $\sim$1300~K.   

If the inner rim of the dust component of the disk is determined by
the sublimation temperature of the dust, this puts
constraints on the type of dust in the disk. For instance, this
excludes dust species with high dust condensation temperatures such as
e.g.~metallic iron or alumina.  Instead, we should expect the dust to
be composed primarily of dust with a sublimation temperature close to
1300~K.  

\subsection{The size of the gas disk}

We can use our measurements to estimate the (radial) size of the
  gas disk.  If we assume that the overtone absorption originates
from gas in Keplerian rotation in the disk, the gas velocity at a
distance $r$ from the center of mass is
\begin{equation}
v = \sqrt{\frac{GM}{r}} \; ,
\end{equation}
\noindent where $G$ is the gravitational constant, $M$ is the total
mass of the binary system (the stars are much closer to each other
than the disk is to the stars so they can be treated as a point mass).
Since we find $v_{\rm in}$ of $13 \pm 3$ \kms and $v_{\rm
out}$ of $9.2 \pm 0.8$ \kms, we conclude that $R_{\rm out}/R_{\rm
in} = (2.0 \pm 0.5)$, compatible with the results by
\citet{Acke2013}, who determined that the outer radius was $(2.2 \pm
0.3)~{\rm R_{in}}$ based on their model fits to the interferometric
observations and mid-IR SED. 

The value for $R_{\rm in}$ is hard to determine, but is most likely
between 10 AU \citep{Dominik2003} and 15 AU \citep{Acke2013}; this
would then correspond to outer radii of $20\pm 5$ and $30\pm 8$ AU and
disk surface areas of $900\pm 300$ and $2100\pm 700$ AU$^2$
respectively. The lower values compare especially well to what we
determined for the \coo gas in the mid-IR where we found an emitting
area of 1300 AU$^2$ (assuming an inclination angle of 60$^{\circ}$,
\paperi).

Thus, the circumbinary disk of HR 4049 is a radially extended,
  gas-rich disk. However, it is not a flat disk. Indeed, the high
  temperatures of the gas ensure that hydrostatic pressure will still
  hold up the disk to appreciable vertical scale heights as well, but
  less than the $H/R \ge 1/3$ that was determined by \cite{Dominik2003}.

\subsection{On the Mass of HR~4049}

From $v_{\rm in}$ of the disk we are able to estimate a mass of the
binary system. Using an inclination angle of 60$^\circ$, we determine
a deprojected $v_{\rm in}$ of 15 $\pm$ 4 \kms at the inner radius;
since this is the FWHM velocity, we find a tangential velocity of
$\sim7.5$ \kms for the gas in Keplerian rotation at the inner
radius. Adopting an inner radius of 15~$\pm 1$~AU, we find a total
mass of $0.94 \pm 0.18$ M$_\sun$ for the binary. If we then apply the
mass function for the primary determined by \citet[$f(M_1)$ = 0.158
  $\pm$ 0.004]{Bakker1998}, we calculate a mass of $0.34 \pm 0.06$
M$_\sun$ for the primary and $0.60 \pm 0.11$ M$_\sun$ for the
secondary. If we use an inner radius of 10~AU, we determine smaller
masses for the binary and individual stars ($M_{\star\star} =
0.64~{\rm M}_\sun, M_1 = 0.17~{\rm M}_\sun, M_2 = 0.46~{\rm M}_\sun$).

The mass we find for the primary at an inner radius of 15~AU is
close to the $0.4 \pm 0.1$~M$_\sun$ estimated by
\citet{Acke2013}. It is thus interesting to 
contemplate the implications of such a low mass for the evolution
of the primary. When a star is on the red giant branch (RGB), it burns
hydrogen in a shell and adds helium to the core. For a low-mass star,
with a degenerate core, the temperature for helium ignition is
attained when the core mass reaches 0.45~M$_\sun$. Unless part of the
core was removed during the common envelope phase, this suggests that
either HR~4049 never ignited helium in its core or that it had a
non-degenerate core on the RGB, which can ignite helium at lower masses. However,
this latter scenario would require a much higher initial mass for HR 4049,
and would suggest that it has lost a great deal of material
which is currently missing from the system. 

When low-mass stars terminate their evolution without igniting helium
in their cores, they will end their lives as helium white dwarfs.
However, HR 4049 is still a giant star. This suggests that
something else may be occurring in this system. Theoretical
predictions show that hydrogen shell flashes occur on helium white
dwarfs which have sufficient hydrogen atmospheres remaining; these
flashes are also thought to return the white dwarf to giant sizes for
very short periods of time \citep[$\sim$\ten{3} yr,][]{Althaus2001}.
The short duration of these flashes is inconsistent with the
observational history of HR 4049. Perhaps it has lost most of its
envelope, but has kept enough for it to sustain a longer-term hydrogen
burning shell. To our knowledge, no evolutionary models of such an
object have been attempted.  

\section{Conclusion}
\label{sec:conclusion}

Based on our new analysis and re-interpretation of near-IR
high-resolution spectoscopic data of HR~4049, we find that the
circumbinary disk surrounding the system is hot and radially extended,
consistent with the results from the mid-IR observations described in
\paperi and by the interferometry described by \citet{Acke2013}. We
find evidence that the fundamental CO emission originates from within
the central, dust-free cavity of the disk. In the absence of dust (and
thus ignoring radiation pressure), viscous dissipation will cause this
gas to end up accreting onto one of the stars in the binary system.
The gas in the circumbinary disk must play a significant role in
determining the physical properties and geometry of the disk. Future
radiative transfer modeling should elucidate these effects.

The mass of the system and the individual components are both very low
and consistent with mass loss. In particular, the total mass of the
system would correspond to a main sequence lifetime on the order of 10
Gyr; such an age is incompatible with the abundances of the
system. Therefore, the system must have lost a considerable amount of
mass during its evolution.

\acknowledgements
We acknowledge the support from the Natural Sciences and Engineering
Research Council of Canada (NSERC).
We would like to thank Ken Hinkle and Sean Brittain for their help in
providing us with the Gemini-Phoenix data.
This research has also made use of NASA's Astrophysics Data System
Bibliographic Services and the SIMBAD database, operated at CDS,
Strasbourg, France.


\begin{thebibliography}{}

\bibitem[{Acke} et~al.(2013){Acke}, {Degroote}, {Lombaert}, et~al.]{Acke2013}
{Acke} B., {Degroote} P., {Lombaert} R., et~al., 2013, {\em \aap\/} 551, A76

\bibitem[{Althaus} et~al.(2001){Althaus}, {Serenelli}, \&
  {Benvenuto}]{Althaus2001}
{Althaus} L.G., {Serenelli} A.M., {Benvenuto} O.G., 2001, {\em \mnras\/} 323,
  471

\bibitem[{Bakker} et~al.(1998){Bakker}, {Lambert}, {Van Winckel},
  et~al.]{Bakker1998}
{Bakker} E.J., {Lambert} D.L., {Van Winckel} H., et~al., 1998, {\em \aap\/}
  336, 263

\bibitem[{Cami} et~al.(2010){Cami}, {van Malderen}, \& {Markwick}]{Cami2010a}
{Cami} J., {van Malderen} R., {Markwick} A.J., 2010, {\em \apjs\/} 187, 409

\bibitem[{Cami} \& {Yamamura}(2001){Cami} \& {Yamamura}]{Cami2001}
{Cami} J., {Yamamura} I., 2001, {\em \aap\/} 367, L1

\bibitem[{Dominik} et~al.(2003){Dominik}, {Dullemond}, {Cami}, \& {van
  Winckel}]{Dominik2003}
{Dominik} C., {Dullemond} C.P., {Cami} J., {van Winckel} H., 2003, {\em \aap\/}
  397, 595

\bibitem[{Goorvitch}(1994){Goorvitch}]{Goorvitch1994}
{Goorvitch} D., 1994, {\em \apjs\/} 95, 535

\bibitem[{Hinkle} et~al.(2003){Hinkle}, {Blum}, {Joyce}, et~al.]{Hinkle2003}
{Hinkle} K.H., {Blum} R.D., {Joyce} R.R., et~al., 2003, In: {Guhathakurta} P.
  (ed.), {\em Society of Photo-Optical Instrumentation Engineers (SPIE)
  Conference Series\/}, vol. 4834 of {\em Society of Photo-Optical
  Instrumentation Engineers (SPIE) Conference Series\/}, pp. 353--363

\bibitem[{Hinkle} et~al.(2007){Hinkle}, {Brittain}, \& {Lambert}]{Hinkle2007}
{Hinkle} K.H., {Brittain} S.D., {Lambert} D.L., 2007, {\em \apj\/} 664, 501

\bibitem[{Lambert} et~al.(1988){Lambert}, {Hinkle}, \& {Luck}]{Lambert1988}
{Lambert} D.L., {Hinkle} K.H., {Luck} R.E., 1988, {\em \apj\/} 333, 917

\bibitem[{Lamers} et~al.(1986){Lamers}, {Waters}, {Garmany}, {Perez}, \&
  {Waelkens}]{Lamers1986}
{Lamers} H.J.G.L.M., {Waters} L.B.F.M., {Garmany} C.D., {Perez} M.R.,
  {Waelkens} C., 1986, {\em \aap\/} 154, L20

\bibitem[{Malek} \& {Cami}(2014){Malek} \& {Cami}]{Malek2013}
{Malek} S.E., {Cami} J., 2014, {\em \apj\/} 780, 41

\bibitem[{Mathis} \& {Lamers}(1992){Mathis} \& {Lamers}]{Mathis1992}
{Mathis} J.S., {Lamers} H.J.G.L.M., 1992, {\em \aap\/} 259, L39

\bibitem[{Najita} et~al.(1996){Najita}, {Carr}, {Glassgold}, {Shu}, \&
  {Tokunaga}]{Najita1996}
{Najita} J., {Carr} J.S., {Glassgold} A.E., {Shu} F.H., {Tokunaga} A.T., 1996,
  {\em \apj\/} 462, 919

\bibitem[{Partridge} \& {Schwenke}(1997){Partridge} \&
  {Schwenke}]{Partridge1997}
{Partridge} H., {Schwenke} D., 1997, {\em "Journal of Chemical Physics"\/} 106,
  4618

\bibitem[Rothman et~al.(2013)Rothman, Gordon, Babikov, et~al.]{Rothman20134}
Rothman L., Gordon I., Babikov Y., et~al., 2013, {\em Journal of Quantitative
  Spectroscopy and Radiative Transfer\/} 130, 4 , \{HITRAN2012\} special issue

\bibitem[{Spitzer}(1978){Spitzer}]{Spitzer1978}
{Spitzer} L., 1978, {Physical processes in the interstellar medium}. New York
  Wiley-Interscience

\bibitem[{Van Winckel} et~al.(1995){Van Winckel}, {Waelkens}, \&
  {Waters}]{VanWinckel1995}
{Van Winckel} H., {Waelkens} C., {Waters} L.B.F.M., 1995, {\em \aap\/} 293, L25

\bibitem[{Waelkens} et~al.(1991){Waelkens}, {Lamers}, {Waters},
  et~al.]{Waelkens1991a}
{Waelkens} C., {Lamers} H.J.G.L.M., {Waters} L.B.F.M., et~al., 1991, {\em
  \aap\/} 242, 433

\bibitem[{Waters} et~al.(1992){Waters}, {Trams}, \& {Waelkens}]{Waters1992}
{Waters} L.B.F.M., {Trams} N.R., {Waelkens} C., 1992, {\em \aap\/} 262, L37

\end{thebibliography}
%

\end{document}